\documentclass[aip,twocolumn,letterpaper]{revtex4}

\usepackage{fullpage}

\usepackage{xcolor}

\usepackage{graphics}
\usepackage{epsfig}

\begin{document}
\title{Changed paradigm of fast magnetic reconnection}
\author{Allen H. Boozer}
\affiliation{Columbia University, New York, NY  10027, ahb17@columbia.edu}

\begin{abstract}

Although magnetic reconnection takes place in three-dimensional space, reconnection theory has focused on two-dimensional models for more than sixty years.  Well-posed three-dimensional mathematics associated with the theory of fluid mixing provides a predictive and compelling explanation for why fast magnetic reconnection is  prevalent---exponentially large variations in the separations between magnetic field lines.  The proofs have been simplified to remove any rational reason to maintain a focus on two dimensional models, which fail to represent the mathematical properties of three-dimensional space.

\end{abstract}

\date{\today} 
\maketitle



Fast magnetic reconnection is traditionally based on two-dimensional models \cite{Zweibel:review,Loureiro:2016}.  Such models do not represent the three-dimensional evolution in a well-posed problem with (i) credible boundary conditions and (ii) a drive for the evolution. 

\begin{figure}
\centerline { \includegraphics[width=3.0in]{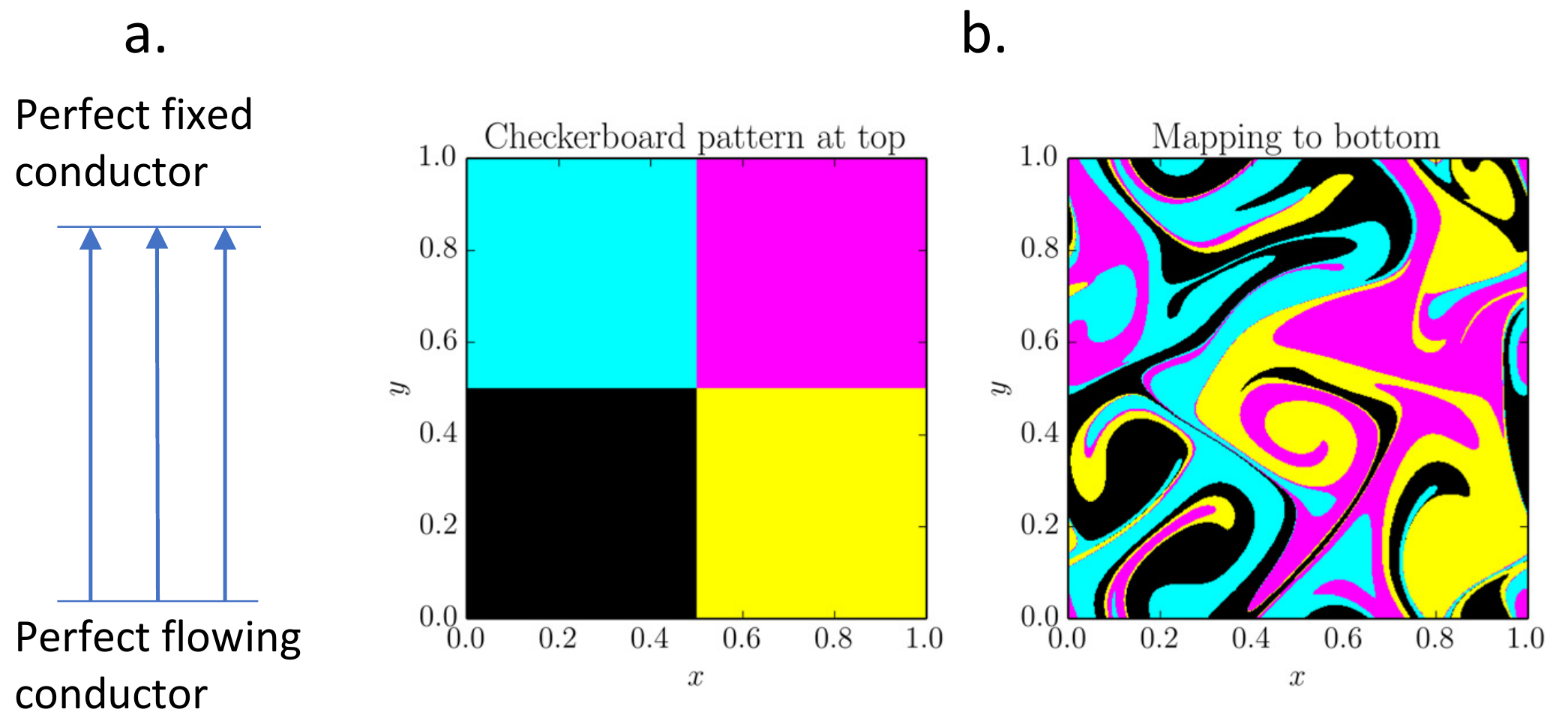} }
\caption{An ideal magnetic evolution is driven by initially straight lines of $\vec{B}$ intersecting perfect conductors, a  flowing conductor at $z=0$ and a  fixed conductor at $z=L$.  The distortion of square flux  tubes by a periodic flow,  Figure \ref{fig:reconn-model}b, was calculated by Yi-Min Huang and used in A. H. Boozer, Nucl. Fusion \textbf{55}, 025001 (2015). }
\label{fig:reconn-model}
\end{figure}

For a well-posed problem \cite{Boozer:prevalence}, consider an initial $z$-directed magnetic field that intercepts two perfectly-conducting planes, Figure \ref{fig:reconn-model}a.  The $z=0$ surface moves with a divergence-free flow $\vec{w}=\hat{z}\times\vec{\nabla} H(x,y,t)$ while the $z=L$ surface is rigid.  The plasma between the planes is highly conducting---initially perfectly conducting---so   \cite{Newcomb}  the magnetic field lines are unbroken and move with a velocity $\vec{u}$;
\begin{equation}
\frac{\partial\vec{B}}{\partial t} = \vec{\nabla}\times(\vec{u}\times\vec{B}). \label{ideal}
\end{equation}
Each magnetic field line has the initial coordinates $(x_0,y_0)$ for all $z$.  At $z=0$, $dx/dt=-\partial H/\partial y$ and $dy/dt=\partial H/\partial x$.  At $z=L$, $x=x_0$ and $y=y_0$ for all $t$.  The evolution time is given by the typical gradient in $\vec{w}$, $\tau_{ev}=1/||\vec{\nabla}\vec{w}||$.

The evolution makes the lines of $\vec{B}$ ever more complicated until line-breaking effects, such as resistivity, become competitive.  To be competitive, resistive effects must be enhanced by the magnetic Reynolds number $R_m\equiv \tau_\eta/\tau_{ev}$, where $\tau_\eta=\mu_0a^2/\eta$ and $a$ is a characteristic spatial scale. For example, in the solar corona \cite{Zweibel:review} $R_m\sim10^{12}$.  Without justification, two-dimensional simulations are started with a current sheet in which the current density is $R_m$ times larger than the characteristic density $j_c\equiv B/\mu_0a$.

Many stream functions $H(x,y,t)$ are physically reasonable \cite{Boozer:null}, but except for special cases, neighboring streamlines in the $z=0$ surface separate exponentially as in Figure \ref{fig:reconn-model}b.  As will be shown, the evolution of the magnetic field implies the ideal solution fails on the time scale $\sim\tau_{ev}\ln(R_m)$.

An understanding of magnetic field line breaking requires an Ohm's law.  A standard form in a plasma moving with a velocity $\vec{v}$ is 
\begin{eqnarray}
&& \vec{E} + \vec{v}\times\vec{B} = \left(\frac{c}{\omega_{pe}}\right)^2 \mu_0\frac{\partial \vec{j}}{\partial t} +\vec{\mathcal{R}}. \label{E1} 
\end{eqnarray}
The $\partial\vec{j}/\partial t$ term is due to the inertia of the lightest current-carrying particle, which is the electron; $\omega_{pe}\equiv \sqrt{ne^2/\epsilon_0 m_e}$, where $n$ is the number density of electrons of mass $m_e$.  The magnetic evolution $\partial \vec{B}/\partial t=-\vec{\nabla}\times\vec{E}$ is simplified by defining an effective magnetic field
\begin{eqnarray}
&& \vec{\mathcal{B}} \equiv \vec{B} + \vec{\nabla}\times \left( \left(\frac{c}{\omega_{pe}}\right)^2\vec{\nabla}\times \vec{B} \right) \label{eff.B}
\end{eqnarray}
using $\vec{\nabla}\times\vec{B}=\mu_0\vec{j}$.  Ampere's law ignores the displacement current, but if it were kept, $\vec{\nabla}\times \vec{B}\rightarrow \vec{\nabla}\times \vec{B}-(1/c^2)\partial \vec{E}/\partial t$, a small correction indeed.

Using the effective magnetic field $\vec{\mathcal{B}}$, Equation (\ref{E1}) can be rewritten as
\begin{eqnarray}
&& \vec{E} + \vec{u}\times\vec{\mathcal{B}} = \left(\frac{c}{\omega_{pe}}\right)^2 \mu_0\frac{\partial \vec{j}}{\partial t} -\vec{\nabla}\Phi+\mathcal{E}_{ni}\vec{\nabla}\ell.  \label{E2} \hspace{0.2in}
\end{eqnarray}
 The $\vec{\mathcal{B}}\times$ components of Equation (\ref{E1})  are balanced by $\vec{u}\times\vec{\mathcal{B}}$, which defines the velocity $\vec{u}$.  The $\vec{\mathcal{B}}\cdot$ components can in large part be balanced by $\vec{\mathcal{B}}\cdot\vec{\nabla}\Phi$, but $\Phi$ must be a well-behaved, single-valued potential.  The non-ideal electric field $\mathcal{E}_{ni}\vec{\nabla}\ell$, where $\ell$ is the distance along an effective magnetic field line, is introduced to make this possible.  

$\mathcal{E}_{ni}$ is constant along the effective magnetic field lines and is chosen to obtain the correct conditions at boundaries and null points \cite{Boozer:null} or for the loop voltage in a torus. 

The evolution equation for the effective magnetic field is
\begin{eqnarray}
&& \frac{\partial \vec{\mathcal{B}} }{\partial t} =\vec{\nabla}\left(\vec{u}\times\vec{\mathcal{B}} -\mathcal{E}_{ni}\vec{\nabla}\ell \right).  \label{eff B ev}
\end{eqnarray}
When $\vec{\mathcal{B}}(\vec{x},t)$ is known, Equation \ref{eff.B} can be solved for the true magnetic field $\vec{B}(\vec{x},t)$.

Equation (\ref{ideal}) for an ideal evolution is broken by both the finiteness of the electron skin depth $c/\omega_{pe}$ and the non-ideal part of the electric field $\mathcal{E}_{ni}$.  Both quantities are extremely small in both natural and laboratory plasmas of practical interest.  The $c/\omega_{pe}$ term in Equation (\ref{eff.B})  is diffusive,  $\vec{\nabla}\times\vec{\nabla}\times\vec{B} =-\nabla^2\vec{B}$.  As shown in \cite{Boozer:null}, two magnetic field lines that come within $c/\omega_{pe}$ at any point along their trajectories become indistinguishable in an evolution.  Reconnection can freely occur on the $c/\omega_{pe}$ scale, so this distance must be extremely small compared to a typical scale $a$ for an ideal evolution to be a useful approximation.  In the solar corona, $c/\omega_{pe}\sim10^{-9}a$.

While estimating the magnitude of $\mathcal{E}_{ni}$, one can ignore the effect of $c/\omega_{pe}$.  The non-ideal electric field $\mathcal{E}_{ni}\approx \eta \bar{j}$, where $\bar{j}$ is an average of the current density along the magnetic field lines and $\eta$ is the resistivity.  The characteristic value of $\mathcal{E}_{ni}$ is $\eta B/\mu_0a$.  The magnetic Reynolds number is $R_m = \left| \vec{u}\times\vec{B}\right|/(\eta B/\mu_0a)$, which is consistent with $R_m\equiv\tau_{\eta}/\tau_{ev}$.  Typical values of $R_m$ are given in \cite{Zweibel:review}.  


 An analytic expression for the growth of the non-ideal part of $\vec{\mathcal{B}}$ can be obtained while the non-ideal part is small.  Write the effective magnetic field in the Clebsch form \cite{Stern:1970}, $\vec{\mathcal{B}}=\vec{\nabla}\alpha\times\vec{\nabla}\beta$.  Since $\mathcal{E}_{ni}$ is constant along the lines of $\vec{\mathcal{B}}$, the effective electric field has the functional form $\mathcal{E}_{ni}(\alpha,\beta,t)$.  Equation (\ref{eff B ev}) and a direct differentiation of $\vec{\mathcal{B}}=\vec{\nabla}\alpha\times\vec{\nabla}\beta$, 
\begin{eqnarray}
\frac{\partial  \vec{\mathcal{B}} }{\partial t} &=&\vec{\nabla}\times\left(\frac{\partial\alpha}{\partial t}\vec{\nabla}\beta-\frac{\partial\beta}{\partial t}\vec{\nabla}\alpha -\vec{\nabla}g\right),
\end{eqnarray}
give $\partial\mathcal{B}/\partial t$ in two forms, which can be equated.  Using $\vec{u}\times\vec{\mathcal{B}}=(\vec{u}\cdot\vec{\nabla}\beta) \vec{\nabla}\alpha - (\vec{u}\cdot\vec{\nabla}\alpha) \vec{\nabla}\beta$, the result is three scalar equations
\begin{eqnarray}
\frac{\partial\alpha}{\partial t}+ \vec{u}\cdot\vec{\nabla}\alpha &=&\frac{\partial g}{\partial\beta}; \hspace{0.2in} \frac{\partial\beta}{\partial t}+ \vec{u}\cdot\vec{\nabla}\beta =-\frac{\partial g}{\partial\alpha}; \hspace{0.2in} \\
\frac{\partial g}{\partial\ell} &=& \mathcal{E}_{ni}(\alpha,\beta,t).
\end{eqnarray}
The gauge function $g$ can depend on all three spatial coordinates and time, $g(\alpha,\beta,\ell,t)$.

Positions in ordinary Cartesian coordinates are given as functions of the Clebsch coordinates by $\vec{x}(\alpha,\beta,\ell,t)=x(\alpha,\beta,\ell,t)\hat{x} +  y(\alpha,\beta,\ell,t)\hat{y} + z(\alpha,\beta,\ell,t)\hat{z}.$  The notation $\partial/\partial t$ implies the Cartesian coordinates are held fixed while $(\partial/\partial t)_c$ implies the Clebesch coordinates are held fixed;
\begin{eqnarray}
0 &=& \frac{ \partial \vec{x}}{\partial t} \\
&=& \left(\frac{ \partial \vec{x}}{\partial t} \right)_c + \frac{\partial \vec{x}}{\partial \alpha} \frac{\partial\alpha}{\partial t} + \frac{\partial \vec{x}}{\partial \beta} \frac{\partial\beta}{\partial t} + \frac{\partial \vec{x}}{\partial \ell} \frac{\partial\ell}{\partial t};\hspace{0.2in}\\
\vec{u} &=& \frac{\partial \vec{x} }{\partial\alpha} \vec{u}\cdot\vec{\nabla}\alpha +\frac{\partial \vec{x} }{\partial\beta} \vec{u}\cdot\vec{\nabla}\beta+\frac{\partial \vec{x} }{\partial\ell} \vec{u}\cdot\vec{\nabla}\ell
\end{eqnarray}
for an arbitrary vector $\vec{u}$.  The velocity of the Clebsch coordinates through the Cartesian coordinates is 
\begin{eqnarray}
 \left(\frac{ \partial \vec{x}}{\partial t} \right)_c &=& \vec{u} - \frac{\partial \vec{x}}{\partial \alpha}\frac{\partial g}{\partial \beta} +\frac{\partial \vec{x}}{\partial \beta}\frac{\partial g}{\partial \alpha} \nonumber \\
 &&- \left(\frac{\partial \ell}{\partial t} + \vec{u}\cdot\vec{\nabla}\ell\right)\frac{\partial\vec{x}}{\partial\ell}.
\end{eqnarray}
Since $\vec{u}\cdot\vec{\mathcal{B}}$ is arbitrary, the choice $\partial\ell/\partial t + \vec{u}\cdot\vec{\nabla}\ell=0$ can and will be made. 

$\vec{\mathcal{B}}$ evolves  ideally when $\mathcal{E}_{ni}=0$, and the gauge $g(\alpha,\beta,\ell,t)$ can be chosen to be zero.  The Clebsch coordinates that have this feature $(\alpha_I,\beta_I,\ell)$ are called ideal.  The resulting equation $(\partial\vec{x}/\partial t)_c=\vec{u}$ implies $\vec{x}(\alpha_I,\beta_I,\ell,t)$ is the transformation to the Lagrangian coordinates of the field line flow.  Not surprisingly, the Clebsch coordinates of an ideally evolving $\vec{\mathcal{B}}$  move with the flow of the lines of $\vec{\mathcal{B}}$.

While non-ideal effects are small, the effective magnetic field can be taken to be an ideally evolving field plus a non-ideal field, $\vec{\mathcal{B}} = \vec{\mathcal{B}}_I +\vec{\mathcal{B}}_{ni}$.  Retaining only the first order deviation from an ideal evolution, the Clebsch coordinates obey
\begin{eqnarray}
\alpha &=& \alpha_I - \frac{\partial \mathcal{A}_{ni}\ell}{\partial\beta_I}; \hspace{0.2in}\beta = \beta_I + \frac{\partial \mathcal{A}_{ni}\ell}{\partial\alpha_I};\\
\mathcal{A}_{ni} &\equiv& - \int_0^t \mathcal{E}_{ni}dt, \hspace{0.1in}\mbox{and}\hspace{0.1in} \vec{\mathcal{B}}_{ni} = \vec{\nabla}\mathcal{A}_{ni}\times\vec{\nabla}\ell. \hspace{0.2in} \label{B_ni}
\end{eqnarray}
$\mathcal{A}_{ni}(\alpha_I,\beta_I,t)\vec{\nabla}\ell$ is the non-ideal part of the vector potential, and $\vec{\mathcal{B}}_I=\vec{\nabla}\alpha_I\times\vec{\nabla}\beta_I$.

The ideal Clebsch coordinates $\vec{x}(\alpha_I,\beta_I,\ell,t)$ are the Lagrangian coordinates $(\partial\vec{x}/\partial t)_c=\vec{u}$ of the flow of the effective field $\vec{\mathcal{B}}$, and $\vec{x}_0\equiv\vec{x}(\alpha_I,\beta_I,\ell,t=0)$.  The Jacobian matrix of the transformation from the initial $\vec{x}_0$ to the time $t$ coordinates $\vec{x}(\alpha_I,\beta_I,\ell,t)$ and the gradient of an arbitrary function $f(\alpha_I,\beta_I,\ell)$ are
\begin{eqnarray}
\tensor{J}&\equiv& \left(\begin{array}{ccc}\frac{\partial x}{\partial x_0} &\frac{\partial x}{\partial y_0} &\frac{\partial x}{\partial z_0} \\ \frac{\partial y}{\partial x_0} &\frac{\partial y}{\partial y_0} &\frac{\partial y}{\partial z_0}  \\ \frac{\partial z}{\partial x_0} &\frac{\partial z}{\partial y_0} &\frac{\partial z}{\partial z_0} \end{array}\right)
= \tensor{U}\cdot\left(\begin{array}{ccc}\Lambda_u & 0 & 0 \\0 & \Lambda_m & 0 \\0 & 0 & \Lambda_s\end{array}\right)\cdot\tensor{V}^\dag \label{SVD} \nonumber\\
&=& \hat{U}\Lambda_u\hat{u}+ \hat{M}\Lambda_m\hat{m}+ \hat{S}\Lambda_s\hat{s} \hspace{0.2in}  \mbox{and    } \label{comp SVD}\\
\vec{\nabla}f &=& \left(\tensor{J}^{-1}\right)^\dag\cdot\vec{\nabla}_0f \nonumber\\
&=& \frac{ \hat{u}\cdot \vec{\nabla}_0f}{\Lambda_u}\hat{U} +\frac{ \hat{m}\cdot \vec{\nabla}_0f}{\Lambda_m}\hat{M}+\frac{ \hat{s}\cdot \vec{\nabla}_0f}{\Lambda_s}\hat{S}. \label{grad f}
\end{eqnarray}
Equation (\ref{SVD}) is the Singular Value Decomposition (SVD) of $\tensor{J}$.  In almost all natural flows, the largest singular value $\Lambda_u$  increases exponentially in time, the smallest $\Lambda_s$ decreases exponentially, and the middle singular value $\Lambda_m$ is slowly varying.   See Figure \ref{fig:reconn-model}b for an illustration and \cite{Boozer:null} for examples.  The unit vectors $\hat{U},\hat{M},\hat{S}$ in $x,y,z$  space are the orthonormal eigenvectors of $\tensor{U}$, and $\hat{u},\hat{m},\hat{s}$ in the $x_0,y_0,z_0$ space are the orthonormal eigenvectors of $\tensor{V}$. The Jacobian or determinant of the Jacobian matrix is $J= \Lambda_u  \Lambda_m  \Lambda_s$, which is related to the divergence of the field line flow $(\partial J/\partial t)_c=J \vec{\nabla}\cdot \vec{u}$.   Typically, $\Lambda_u=e^{\gamma_u t}$, where $\gamma_u \sim 1/\tau_{ev}$ is the Lyapunov exponent of the flow.

The ideal evolution of the effective magnetic field can be written as \cite{Stern:1966,Zweibel:review}, $ \vec{\mathcal{B}}_I(\vec{x},t) = (\tensor{J}/J)\cdot \vec{\mathcal{B}}_0$, where $\vec{\mathcal{B}}_0$ is the effective magnetic field at $t=0$, so
\begin{eqnarray}
&& \mathcal{B}_I^2 = \left( \frac{\hat{u}\cdot\vec{\mathcal{B}}_0}{\Lambda_m\Lambda_s}\right)^2 +  \left( \frac{\hat{m}\cdot\vec{\mathcal{B}}_0}{\Lambda_u\Lambda_s}\right)^2+ \left( \frac{\hat{s}\cdot\vec{\mathcal{B}}_0}{\Lambda_u\Lambda_m}\right)^2, \hspace{0.2in}
\end{eqnarray}
The term in $\mathcal{B}_I^2 $ proportional to $(\hat{u}\cdot\vec{\mathcal{B}}_0)^2$ goes to infinity exponentially in time.   The term  proportional to $(\hat{s}\cdot\vec{\mathcal{B}}_0)^2$ goes to zero exponentially.   A bounded magnetic field strength is only possible for a time long compared to $\tau_{ev}$ when the effective magnetic field points in the $\hat{M}$ direction, $\vec{\mathcal{B}}_I(\vec{x},t) \rightarrow \left(\hat{m}\cdot\vec{\mathcal{B}}_0/\Lambda_u\Lambda_s\right)\hat{M}$.  The forces associated with the magnetic field will constrain the flow velocity $\vec{u}$ to ensure this happens. This is no more obscure than how a stream flowing down a mountain twists and turns to follow a gulley.

Equation (\ref{grad f}) for the gradient of a function has the asymptotic form $\vec{\nabla}f \rightarrow \left(\hat{s}\cdot \vec{\nabla}_0f/\Lambda_s\right)\hat{S}$, and becomes exponentially large in the direction in which streamlines of $\vec{u}$ approach each other exponentially; $\Lambda_s$ goes to zero exponentially.  Consequently $\vec{\nabla}\mathcal{A}_{ni} \rightarrow \hat{S}(\hat{s}\cdot\vec{\nabla}_0\mathcal{A}_{ni})/\Lambda_s$. The magnetic field lines become oriented in the $\hat{M}$ direction, so Equation (\ref{grad f}) when applied to $\vec{\nabla}\ell$ implies $\vec{\nabla}\ell\rightarrow \hat{M}/\Lambda_m$ plus a term in the $\hat{S}$ direction. Equation (\ref{B_ni}) then shows that the non-ideal part of the magnetic field grows exponentially in time \cite{Boozer:ideal-ev},
\begin{eqnarray}
\vec{\mathcal{B}}_{ni}&\rightarrow&\frac{\hat{S}\times\hat{M}}{\Lambda_m\Lambda_s} \hat{s}\cdot\vec{\nabla}_0\mathcal{A}_{ni}(\alpha_I,\beta_I,t).
\end{eqnarray}
$\Lambda_s$ approaches zero exponentially.  The non-ideal field is oriented in the direction in which neighboring flow streamlines exponentiate apart; $\hat{S}\times\hat{M}=-\hat{U}$.   

On a time scale, $\sim\tau_{ev}\ln R_m$ the effective magnetic field will enter a state of fast magnetic reconnection.  On a time scale $\sim\tau_{ev}\ln (a/(c/\omega_{pe}))$, electron inertia produces a large scale magnetic reconnection.


Reconnection theory from  Sweet \cite{Sweet:1958} and Parker \cite{Parker:1957} in the 1950's to the present \cite{Huang:2019} has focused on two-dimensional models of three-dimensional systems.  Modern two-dimensional models were reviewed in 2016 by Zweibel and Yamada \cite{Zweibel:review} and by Loureiro and Uzdensky \cite{Loureiro:2016}.  Magnetic reconnection in two-dimensional models is fundamentally different than in three-dimensional systems.  In two-dimensions, but not in three, an exponential increase in field strength is required for magnetic field lines to exponentiate apart.  Resistivity can compete with evolution in two dimensions only if the current density becomes nearly singular, $j \approx R_mB/\mu_0 a$, by the formation of a current sheet of thickness $\delta_j\approx a/R_m$.  Standard two-dimensional theory assumes but does not explain the formation of this current sheet and consequently makes no prediction on when an evolving magnetic field reaches a rapidly reconnecting state.  In three-dimensional theory \cite{Boozer:null}, the current density when reconnection occurs is  $\sim\ln(R_m) (B/\mu_0a)$. 

The dominance of two-dimensional thoery has been sufficient that the presence of a near-singular current is often considered a requirement for fast magnetic reconnection.   A magnetic null, $\vec{B}=0$, seems a likely place for a near-singular current density \cite{{Craig:2014},Priest:2016}.  However, authors have not considered the implications of the indistinguishability of magnetic field lines that pass near either a null or an X-point.  Lines within a magnetic flux tube of radius $c/\omega_{pe}$ do not remain distinguishable in an evolution, so distinguishable lines do not come within a distance $(a^2 c/\omega_{pe})^{1/3}$ of a null \cite{Boozer:null}.  In a torus, a similar argument implies distinguishable lines do not come within a distance $\sqrt{a c/\omega_{pe}}$ of an X-point.  

The near-singular current of two dimensional theory dissipates magnetic energy, $\vec{j}\cdot\vec{E}\sim \eta j^2$, on the time scale of the reconnection and the strong electric field  can accelerate particles.  In three-dimensional theory, magnetic energy is transferred into Alfv\'en waves, but the full equations for particle drift motion predict particle acceleration \cite{Boozer:part-acc}.

The reconnection due to magnetic field line separation is related to the enhancement of reconnection by turbulence \cite{Lazarian:2020review,Servido:2014}.  Turbulent flow velocities have a short correlation distance compared to the scale of the reconnecting region, which produces an intrinsically slower process than mixing on the scale of the ideal flow velocity $\vec{u}$.


Fast magnetic reconnection is an application of advection-diffusion theory, which explains why the temperature $T$ in a typical room relaxes in tens of minutes, not several weeks as expected for thermal diffusion in air, $D\approx 2\times10^{-5}~$m$^2$/s.  The mathematical explanation is in the 1984 paper by Hassan Aref \cite{Aref:1984}.  If $D$ were zero, a divergence-free flow of air would move points with fixed $T$ about the room on a time scale $\tau_{ev}=a/v$ without breaking  constant-T contours.  The surface area of constant-T contours would increase exponentially in time as would the rate at which diffusion relaxes temperature gradients; $\tau_d/\tau_{ev}\sim10^4$.  At least two spatial dimensions are required.  In one dimension, an exponential increase in diffusive relaxation would require exponentially large variations in the density of the air.  Divergence-free stirring requires two singular values of the Jacobian matrix, hence two spatial dimensions.  The applicability of advection-diffusion theory to the magnetic field requires three spatial dimensions to avoid an exponential increase in the magnetic field strength.  The mathematics of a three-dimensional magnetic evolution is not represented in a two-dimensional model despite sixty years of effort.  The vorticity, $\vec{\omega}\equiv\vec{\nabla}\times\vec{v}$, in divergence-free fluid mechanics has similar evolution properties.

\section*{Acknowledgements}
This work was supported by the U.S. Department of Energy, Office of Science, Office of Fusion Energy Sciences under Award Numbers DE-FG02-95ER54333, DE-FG02-03ER54696, DE-SC0018424, and DE-SC0019479.


\end{document}